# Mitigating Data Sharing in Public Cloud using Blockchain

**[1]Pratik Patil, [2]Prerna Tulsiani, [3]Dr. Sunil Mane**

[1]Student, [2]Student, [3]Associate Professor

[1,2,3]Department of Computer Science and Engineering,

College of Engineering Pune Technological University, Pune, India

***Abstract:*** Public Cloud Computing has become a fundamental part of modern IT infrastructure as its adoption has transformed the way businesses operate. However, cloud security concerns introduce new risks and challenges related to data protection, sharing, and access control. A synergistic integration of blockchain with the cloud holds immense potential. Blockchain's distributed ledger ensures transparency, immutability, and efficiency as it reduces the reliance on centralized authorities. Motivated by this, our framework proposes a secure data ecosystem in the cloud with the key aspects being Data Rights, Data Sharing, and Data Validation. Also, this approach aims to increase its interoperability and scalability by eliminating the need for data migration. This will ensure that existing public cloud-based systems can easily deploy blockchain enhancing trustworthiness and non-repudiation of cloud data.

## I. INTRODUCTION

Rapidly progressing technology is highly driven by interdisciplinary collaboration. It is fueled by breakthrough advancing research in the field of Big Data, Internet of Things (IoT), Artificial Intelligence, Machine Learning, and many more. The iterative nature of these evolving technologies requires substantial computation power, heavy upfront investments, and on-demand scaling. Cloud technology plays an indispensable role in this contemporary digital landscape. Cloud incorporates specialized tools and frameworks that facilitate quick deployment and testing of live applications, thus offering organizations a competitive edge. The most widely accepted definition of cloud computing technology was given by the National Institute of Standards and Technology [1]: "Cloud computing is a model for enabling ubiquitous, convenient, on-demand network access to a shared pool of configurable computing resources (e.g., networks, servers, storage, applications, and services) that can be rapidly provisioned and released with minimal management effort or service provider interaction."

### 1.1 Decoding Cloud Computing Essentials

Cloud services are globally accessible, and they help improve operational excellence by reducing human interventions. This is achieved through multiple cloud deployment models provided by cloud service providers which cater to the diverse needs of organizations. Cloud deployment refers to the process of configuring and managing resources so that they become easily available over the Internet. This eliminates the need for organizations to maintain physical data infrastructure systems. Following are the three major types of cloud deployment models, which offer services according to client business needs [2]:

***Table 1.1:*** *Cloud Deployment Models*

| Factor | Public Cloud | Hybrid Cloud | Private Cloud |
|---|---|---|---|
| **Auto – Scaling** | High | Moderate | Limited |
| **Security** | Good | Very Secure | Most Secure |
| **Cost** | Low<br>(Pay as you go Model) | Moderate | High<br>(Special Staff Required) |
| **Who is it for?** | Fast growing companies | Good amount of critical data | Banks and Financial Firms |

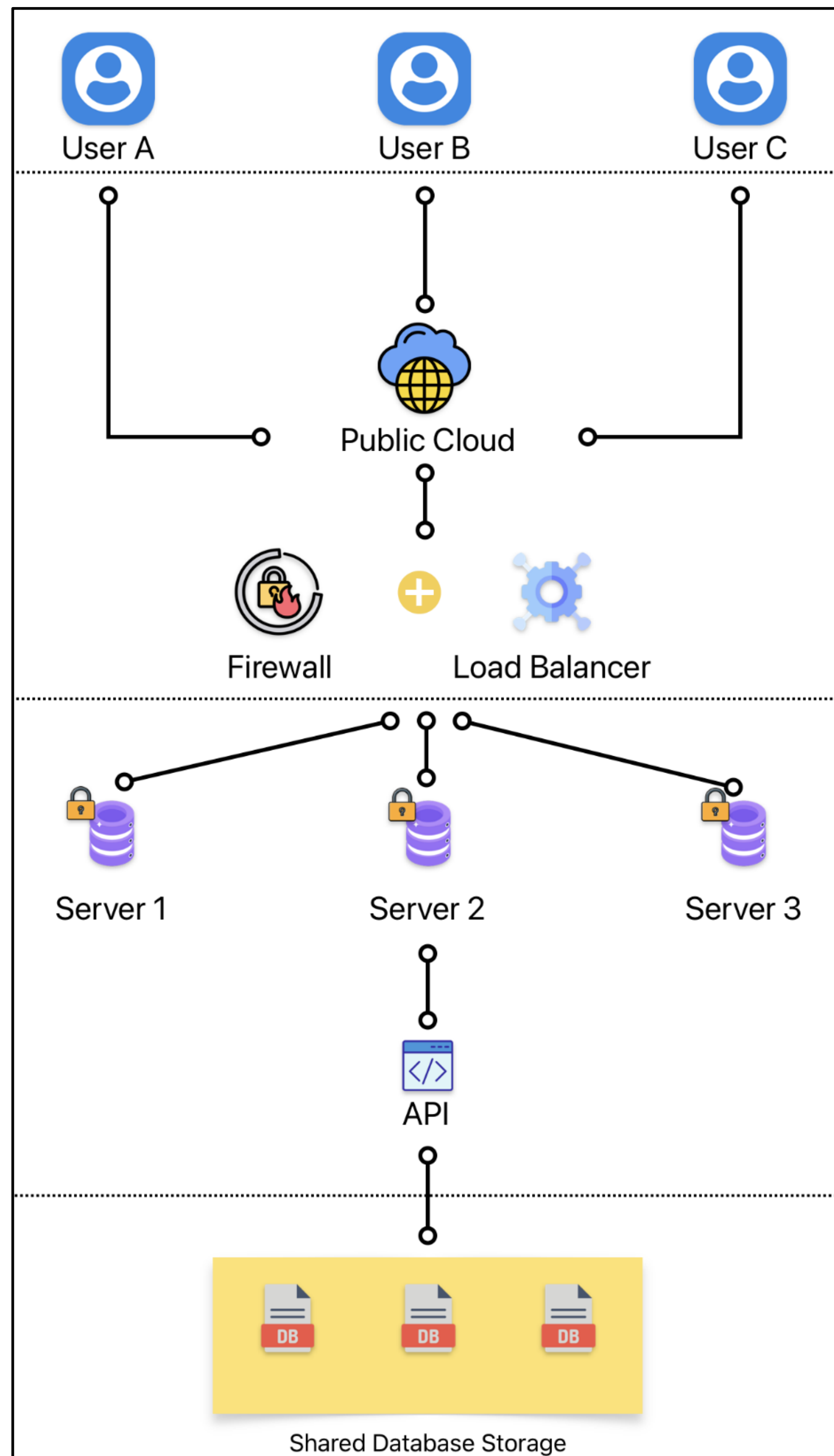

***Figure 1:*** *Cloud Architecture*

A public cloud deployment model benefits its users by making all the resources publicly available over the Internet with a very cost-effective and scalable architecture [3]. Starting with the top-most layer i.e. Application Layer - Variety of applications are hosted in this layer and users interact with the cloud environment leveraging the interactive front-end technologies. This layer helps an organization to highlight the functionality of their system. The layer below the application layer is the Service Provider Layer. This layer encompasses the services delivered by the Cloud Service Providers like the storage facilities, security features, networking strategies, and many more. The next layer is the Processing Layer, which provides virtualization resources, allowing multiple virtual machines to run on a single physical server. APIs (Application Programming Interfaces) act as a bridge between applications and underlying virtual infrastructure. They control the resource sharing programmatically and hence act as a standardized mechanism for seamless communication. At the bottom is the Data Layer which consists of the physical backbone of the cloud architecture. It includes the servers, storage devices, etc. The pooling and resource allocation of this layer is completely managed by the Cloud Service Providers based on the support requested by the processing layer.

## 1.2 Harnessing Benefits of Blockchain

The fundamental shift in how organizations store and process their data is greatly impacted by the public cloud deployment model. Cloud architecture provides a virtualized isolated environment for all the stakeholders. This allows them to run their applications independently. However, the multi-tenancy model involving multiple users or "tenants" sharing the same platform and resources, introduces a lot of security risks, and hence maintaining a balance between robust security measures and resource optimization has become a crucial concern [4]. Therefore, the shared nature of the public cloud introduces many complexities related to data transferring and data protection. Blockchain, a transforming technology is gaining momentum due to its unique approach to managing data and enforcing trust. Its emergence can be traced back to the introduction of Bitcoin in 2008 by Santoshi Nakamoto [5]. Since then, blockchain has gained a lot of attention due to its decentralized, transparent, tamper-resistant, anonymous, and auditability properties.

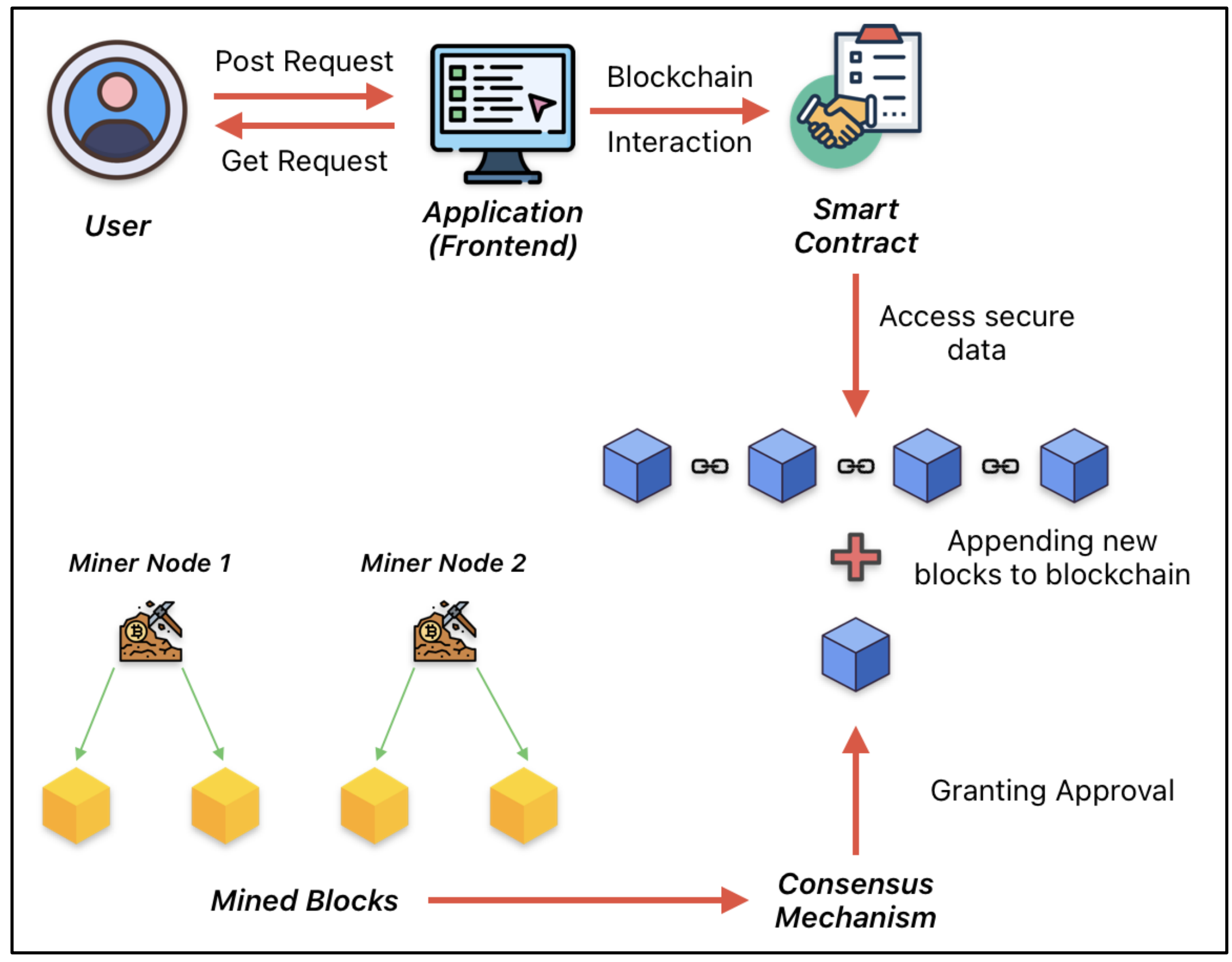

***Figure 2:*** *Blockchain Architecture*

Blockchain is a peer-to-peer network where the distributed participants known as nodes communicate directly with each other. This reduces the need for any intermediaries. Data is stored in blockchain by inter-linking blocks where each block is linked to the previous one through a cryptographic hash forming a chain. A new block is added to blockchain by circulating an agreement protocol among nodes [6]. A self-executing data agreement known as a Smart Contract is deployed on blockchain which enables real-time settlement of data transactions. It acts as a transit between applications and blockchain by embedding predefined terms and programmable logic. This logic gets triggered when specified conditions are satisfied which automates the processes of transferring assets or updating records. To attain the core principle of security i.e. trust, a resolution process is required for the group of nodes so that a decision can be made regardless of any individual choices and this technique to reach a conclusion is known as Consensus Mechanism. In the field of cloud computing, a voting-based consensus mechanism exhibits notable efficiency over proof-based counterparts, as it is resource-effective, scalable and prioritizes efficiency [7].

# II. Related Work

The shared infrastructure of the public cloud environment faces many security vulnerabilities due to dependencies on third-party providers introducing concerns about data privacy. For example, AWS (Amazon Web Services) had 4 hours of downtime on February 28th, 2017, resulting in thousands of websites and applications becoming completely inaccessible [8]. Hence, integrating blockchain with the cloud offers immense scope. Moreover, blockchain's decentralized and tamper-resistant framework complements the cloud's scalable and flexible infrastructure. Leveraging this integrated ecosystem organizations can easily mitigate risks associated with data sharing and strengthen trust among stakeholders.

In [9], authors designed an architecture to store Electronic Health Records in a blockchain-based network. This shows how fast blockchain technology has evolved since its emergence. As it maintains stringent security standards and safeguards sensitive health information. This proposed architecture introduced a new incentive mechanism for the creation of new blocks in the blockchain. Also, their design ensures that each block gives equal contribution in this process for conducting the new block insertion process. Blocks with the least significance are selected in each iteration and after completing the task their significance gets updated and they are given incentive charges.

BigchainDB [10] is a blockchain database that is a combination of blockchain and database supporting a wide range of proof-of-concepts. It allows developers to deploy their applications with high throughput, low latency, decentralized control, and immutable data storage. With user feedback and iterative improvements, the next version of BigchainDB software aims to include the integration of Tendermint for inter-node networking and Byzantine fault tolerant (BFT) consensus.

ProvChain [11] is a blockchain-based data provenance architecture in a cloud environment. This architecture presented a solution to embed metadata of the history of creation and operations performed on a cloud object in a tamper-proof blockchain, thus enhancing user privacy and reliability. However, the implementation of ProvChain for multiple cloud servers may require a lot of computational speed and cost and, hence limit the scalability.

In [12], authors found various security risks in cloud and network issues like compromised credentials, hacked interfaces and APIs, account hijacking, permanent data loss, DoS attacks, and inadequate diligence. Further, they proposed how various aspects of cloud security could be solved with the help of blockchain technology. Thus, making the system more secure by adding data encryption, service-level agreements, cloud data management, and interoperability. However, the increasing size of this distributed ledger may impact the cost and degrade the performance.

Here [13], the authors proposed a novel blockchain-based data preservation system for medical data and demonstrated various performance evaluation results by showing many comparisons. They focused on explaining why they used Ethereum over Bitcoin. The key parameters of comparison were block size, response time, and transactional gas required. Also, their research highlighted the importance of fixed block size. If the block storage capacity is under-utilized this will lead to wastage of resources while, if it is over-utilized then breaking the data into chunks and then storing it will increase the time required to verify the consistency of blockchain.

# III. Research Gap

The public cloud environment is indeed a very vast and complex distributed system architecture. This is because the cloud architecture not only addresses the continuous growth of the industrial ecosystem but also empowers organizations to adapt to digital transformation. This in turn leads to seamless integration of all the dynamic stakeholders. Organizations need to enforce legal compliance regulations to protect financial records, strategic plans, customer confidential information, and intellectual property. Systems accessing cloud services have become sophisticated and involve a diverse range of individuals. Hence, there is a necessity of building a focused approach which will distinguish responsibilities and optimize the productivity of such comprehensive systems.

Based on a comprehensive review of previous research in this field, we classified various frameworks into four distinct types. After performing a detailed analysis to determine whether they have incorporated the core strategies that this framework aims to target.

*Table 3.1: Comparison of Core Strategy Adoption*

| Types | Access Control Policies | Session Tokens | Validation and Verification | Off-Chain and On-Chain Bifurcation |
|---|---|---|---|---|
| **Electronic Health Records Blockchain Based System** | ✓ | ✓ | × | × |
| **ProvChain** | ✓ | ✓ | ✓ | × |
| **BigchainDB** | ✓ | ✓ | × | × |
| **On-chain vs off-chain storage for blockchain integration** | × | × | ✓ | ✓ |

## 3.1 Strategy for Protection and Compliance

Access Control plays a pivotal role in governing how users with different roles will access cloud resources and under which circumstances [14]. The framework stores role-based access control policies in blockchain. Distributed ledgers in blockchain are non-tampered as they are spread across multiple nodes and each participant in the blockchain network maintains an identical copy of the ledger. Each block has a recorded date timestamp and is provided with a unique cryptographic hash signature which guarantees the integrity of the ledger [15]. Specifically, there is a critical need to understand how development of blockchain-based systems can align with legal standards of an organization.

## 3.2 Verifying Trust

Organizations prioritize trusted cloud service providers, crucial for their reputation and public perception. A foundation of trust can only be achieved by establishing a secure flow of data transactions in the cloud environment. Digital Certificates create a chain of trust. They confirm the legitimacy of individuals affirming secure online transactions. A digital certificate is a digital statement issued by a Certifying Authority (CA) that vouches for the identity of the certificate holder and enables parties to communicate securely by creating an encrypted channel [16]. Hence there is no doubt that digital certificates boost data sharing security. However, there are growing concerns surrounding digital certificates, regarding data manipulation and central authority reliance. A gap lies in exploring blockchain's potential to counter these vulnerabilities and enhance certificate management.

## 3.3 Solutions to Mitigate Energy Consumption

As discussed earlier, previous research presented a solution to store entire cloud data on a blockchain network as this immutable ledger offers a lot of security benefits. However, storing entire cloud data on the blockchain also poses numerous hurdles. The energy-intensive storage mechanism of blockchain could lead to very high computing costs and speed of transactions. Firstly, if whole cloud data is stored on blockchain then the entire information will become publicly accessible hampering the confidentiality of whole data. Secondly, variable-sized blocks are employed to accommodate diverse amounts of cloud data. This may give rise to many issues such as transaction delays, data fragmentation, and consensus complexity. Hence, there is a need to explore solutions that reconcile security while mitigating energy-intensive processes and optimize blockchain performance [17].

# IV. PROPOSED METHODOLOGY

Harnessing the wide-range security benefits of blockchain technology, the framework offers a solution to mitigate data sharing and data trust vulnerabilities in multifaceted public cloud environments. This approach blends advanced access control policies, robust digital certificate mechanism, and cost-effective off-chain data storage strategy integrated with on-chain verification and validation. Following is the outline of the proposed architecture describing the sequence of how the cloud data flows via protected blockchain immutable, decentralized, and non-tampered ledger:

*1) Assigning Pre-Defined Access Privileges:*

This step involves categorizing the stakeholders of an organization based on their responsibilities and granting them permissions aligned with their roles. This is done to regulate high compliance standards within the organization. The data administrators will have the authority to make these role-based settings. After data updation from the frontend application, these rules will get saved in the blockchain network.

*2) Authentication:*

After a user logs into the system the credentials are compared. The role assigned to the user will be verified from the access control policies stored in the blockchain network. Next, the logic embedded in the smart contract is executed and the specified conditions are compared. According to the result obtained following actions are performed:

*2.1) Rejection:* If a user attempts to access data for which they are not intended, then the smart contract will deny the request and the system administrator will get alerted.

*2.2) Acceptance:* If a user has authorised access to the requested data, then smart contract will accept this request and the system will trigger the generation of a session token. Till the session token is valid it will be used to authorise a particular user. The session token has attributes containing user's unique identification number, user role and time validity. This token is cryptographically encrypted and will serve as a temporary access credential for the user inside the system.

*3) Logging:*

The acceptance / rejection of a user captured from the smart contract is recorded in the form of logs. These logs are very useful for identifying percentage of legitimate users entering the system, thus enhancing the security breaches. Also logs contribute as a valuable resource for monitoring and auditing purposes.

*4) Off-Chain Data:*

Generation of a session token, indicates the user is legitimate and is completely authenticated. This initiates a successful connection with the cloud servers. Within this connected ecosystem the architecture is connected to both off-chain and on-chain data. Off-chain data typically resides in traditional database systems, that are not part of the blockchain itself. The contents of this off-chain data vary from large multimedia files, sensitive customer information, transaction details, logs, and other proprietary documents. Reference of off-chain data is linked via cryptographic fixed-size hashes. These hashes are then stored on the blockchain. Conventional methods store whole data on the blockchain which is then queried according to the user's request. This increases the response times, transaction gas fees, and latency levels. Thus, the linkage of on-chain and off-chain data helps in compact representation of large existing datasets. It also facilitates trustworthiness by verifying on-chain hash and off-chain cloud data.

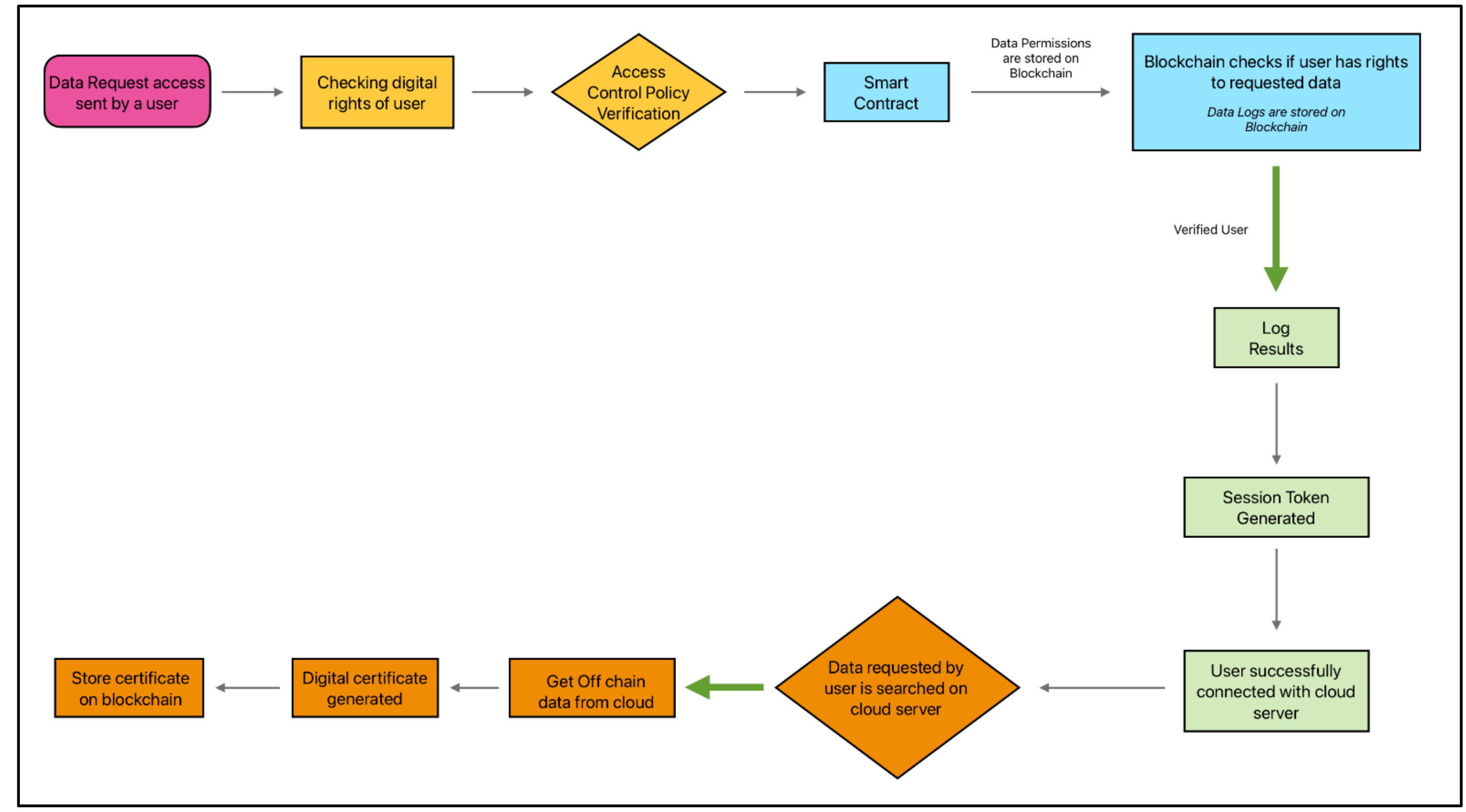

***Figure 3:*** *Proposed System Architecture*

### *5) Digital Certificate:*

In dynamic cloud environments, continuous real-time data transactions occur at changing frequencies. Thus, a non-repudiation mechanism is essential to provide assurance that the involved parties cannot deny the origin of the message or action performed. A digital certificate is a source of irrefutable evidence illustrating the proof of origin and the proof of receipt [18]. The framework creates a digital certificate which comprises of the following essential components:

- Server's unique identifier
- User's unique identifier
- Expiration date
- Fixed-size cryptographic hash generated from off-chain requested data.

## V. ALGORITHM DESIGN

This section focuses on the strategic development of three critical algorithms: the Dynamic Access Control Algorithm for Data Security, the Check if a user has permission to perform a system functionality algorithm, and Blockchain-Based Verification in Off-Chain Storage. These algorithms are meticulously crafted to address key challenges in data security, access control management, and data verification. The design principles behind these algorithms have been highlighted by incorporating examples of real-world scenarios. This showcases the relevance and easy integration of this framework with existing cloud database systems.

The Dynamic Access Control Algorithm (as shown in Algorithm 1) defines a procedure to manage and update access permissions based on individual roles. It categorizes individuals into "Controllers" and "Users". This algorithm grants Controller authority to onboard new individuals to the organization and assign them roles. In addition, controllers can modify access control settings to align with the organization's compliance regulations.

| **Algorithm 1:** Dynamic Access Control Algorithm for Data Security |
|---|
| **INPUTS**:<br>• *userRole*, where userRole ∈ [Controller, User]<br>• *action*, where action ∈ [Onboarding, Assign Role, Create Policy, Update Permission, Check Control, Audit Trail] |

| **OUTPUT**: *validAction*, where validAction ∈ [True, False] | |
|---|---|
| 1: | **function** UPDATE_ACCESS_CONTROL (*userRole*, *action*): |
| 2: | **if** *userRole* = Controller **then** |
| 3: | set *validAction* = True |
| 4: | **if** *action* = Onboarding **then** |
| 5: | // Onboard new user and assign the user a particular role |
| 6: | **else if** *action* = Assign Role **then** |
| 7: | // Assign or update role of a user |
| 8: | **else if** *action* = Create Policy **then** |
| 9: | // Create a policy corresponding to a particular system functionality |
| 10: | **else if** *action* = Update Permission **then** |
| 11: | // Grant or Revoke permission to a user for a particular system functionality |
| 12: | **else if** *action* = Check Control **then** |
| 13: | // View if a user has permission to perform a particular system functionality |
| 14: | **else if** *action* = Audit Trail **then** |
| 15: | // Check past audit trails |
| 16: | **end if** |
| 17: | **else if** *userRole* = user **then** |
| 18: | set *validAction* = False // Since only admin controller can update policies |
| 19: | **end if** |
| 20: | RECORD_AUDIT_TRAIL (*userId*, *action*, *timestamp*, *validAction*) |
| 21: | return *validAction* |

Let's consider an example for instance:

- Admin controller logs into the system and he/she has to onboard a new employee to the organization. This employee will be working in the HR Department. Thus, the employee should be able to access all HR department data.
- From the front end of the application, the controller will navigate to the "Access Control Settings" Page. It will call UPDATE_ACCESS_CONTROL function in the backend. Initially, the function will verify if the userRole corresponds to the Controller role, then it will allow the controller to access other actions. After validating the controller will be able to see these actions: Onboard new users, assign roles, create policy, update permission, check control, and audit trail.
- Now controller will onboard the new employee and assign role to the employee. Let's consider userId = "123".
- The controller will create a new policy named "HR Data Access" and then grant permission to the employee with userId = "123". This will ensure that this new employee can view the HR Department's Data on his/her dashboard.

User Experience plays a significant role as it provides seamless access to authorized system functionalities while restricting access to unauthorized areas. Algorithm 2 automates the process of checking permissions of different users. It ensures that access permissions are granted and revoked appropriately, reducing the risk of compliance violations. Now, continuing the previous example demonstrated earlier in this section, we will try to understand how this framework checks if userId = "123" can access HR Department functionalities.

**Algorithm 2:** Check if a user has permission to perform a system functionality

**INPUTS**:

- *userId*, which is a unique identification number corresponding to a user
- *functionalityName*, where functionalityName is a particular system functionality

**OUTPUT**: *hasPermission*, where hasPermission ∈ [True, False]

| **ASSUMPTION**: “*policiesDictionary*” is a global variable with key: functionalityName & corresponding values: list of userIds who have permission to access that functionality | |
|---|---|
| 1: | **function** CHECK_PERMISSION (*userId*, *functionalityName*): |
| 2: | **if** *userId* $\in$ *policiesDictionary*[*functionalityName*] **then** |
| 3: | set *hasPermission* = True |
| 4: | **else** |
| 5: | set *hasPermission* = False |
| 6: | **end if** |
| 7: | return *hasPermission* |

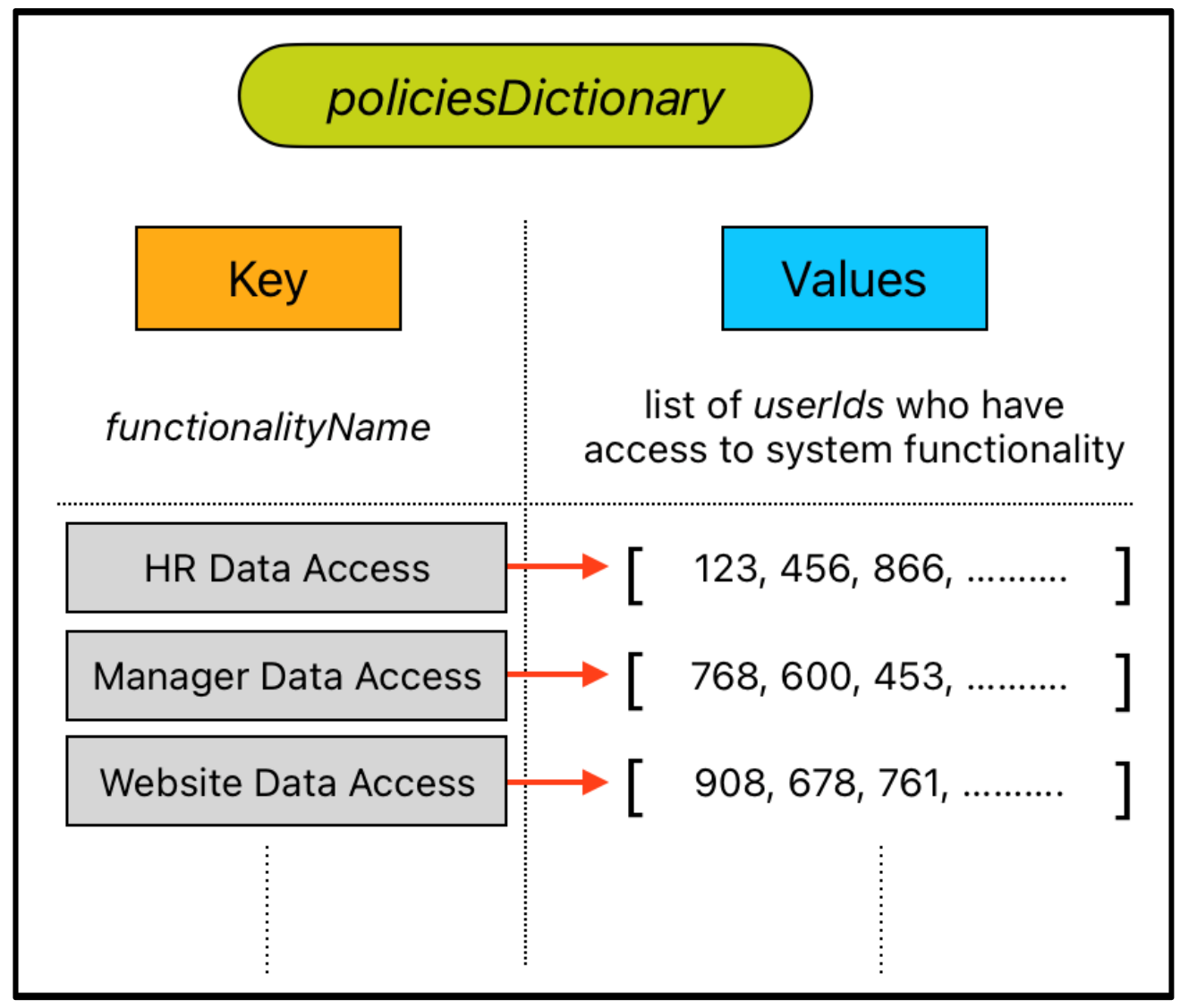

***Figure 4:*** *Access Control Policies Dictionary*

The front end will call CHECK_PERMISSION function with parameters “123” (userId) and “HR Data Access” (functionalityName). The algorithm will search the dictionary for the key “HR Data Access” and then it will check if “123” belongs to the list of userIds. Finally, the algorithm will return user permission status.

The Blockchain-Based Verification for Off-Chain Storage algorithm is designed to bolster integrity and security. This algorithm comprises three classes: BLOCKCHAIN, CLOUD_DATABASE, and LINKER. The BLOCKCHAIN class is responsible for initializing the blockchain structure and adding new blocks to the chain. This structure stores a sequential chain of blocks containing references to off-chain cloud data. On the other hand, the CLOUD_DATABASE class manages the actual storage of data, including file contents and digital certificates. The LINKER class acts as a coordinator, and facilitates the process of linking Off-Chain data (such as file contents and certificates stored in the cloud database) to the On-Chain blockchain. This involves adding blocks to the blockchain that contain relevant information about Off-Chain data, ensuring a tamper-proof verification mechanism.

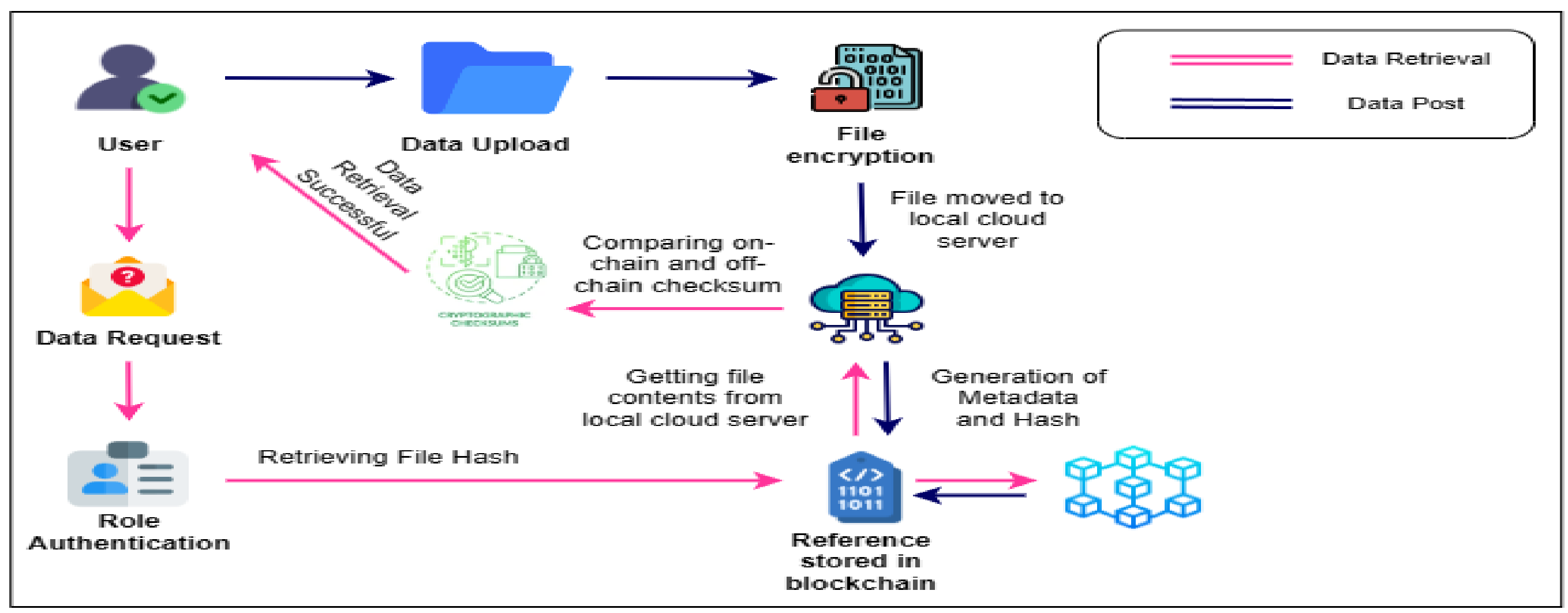

***Figure 5:*** *Data Integrity: On-Chain Hashes and Off-Chain File Retrieval*

**Algorithm 3:** Blockchain-Based Verification for Off-Chain Storage

```
 1:  class BLOCKCHAIN:
 2:      function INIT:
 3:          set chain = []      // creates an empty array for storing chain of blocks
 4:      end function
 5:      function ADD_BLOCK (fileContentHash, certificateId, serverId, userId, expiry):
 6:          chain.append({"fileContentHash" : fileContentHash, "certificateId" : certificateId,
     "serverId": serverId, "userId": userId, "expiry": expiry})
 7:      end function
 8:  end class
 9:  class CLOUD_DATABASE:
10:      function INIT:
11:          set certificates = { }      // creates dictionary for storing certificate linked to a file
12:          set files = { }      // creates a dictionary for storing file contents
13:      end function
14:      function UPLOAD_DATA (fileContentHash, fileContent, certificateId, certificateContent):
15:          set certificates[certificateId] = certificateContent
16:          set files[fileContentHash] = fileContent
17:      end function
18:  end class
19:  class LINKER:
20:      function INIT:
21:          set B = BLOCKCHAIN, set C = CLOUD_DATABASE
22:          B.INIT, C.INIT
23:      end function
24:      function HASH_FILE_CONTENTS (fileContent):
25:          set fileContentHash = HASH (fileContent)
26:      end function
27:      function LINK_OFF_CHAIN_TO_ON_CHAIN:
28:          B.ADD_BLOCK (fileContentHash, certificateId, serverId, userId, expiry)
29:          C.UPLOAD_DATA (fileContentHash, fileContent, certificateId, certificateContent)
30:      end function
31:  end class
```

# VI. IMPLEMENTATION & DISCUSSIONS

The integration of blockchain's distributed ledger with public cloud computing environment provides a robust foundation for secure data storage and data trust.

## 6.1 Addressing Vulnerabilities and Single Points of Failure in Cloud Access Control

The centralized nature of traditional public cloud architecture leads to ineffective access management. Also, there is a possibility of a lot of vulnerabilities being caused due to a single point of failure making it difficult to track user activities. Limited visibility of user permissions across diverse cloud services hampers the privacy of confidential data. The proposed framework, however, stores these access control policies in an immutable ledger which thus ensures that logs remain unaltered and access management becomes easily auditable. This will allow clear visibility of user activities and data ownership, enhancing the accountability of the system.

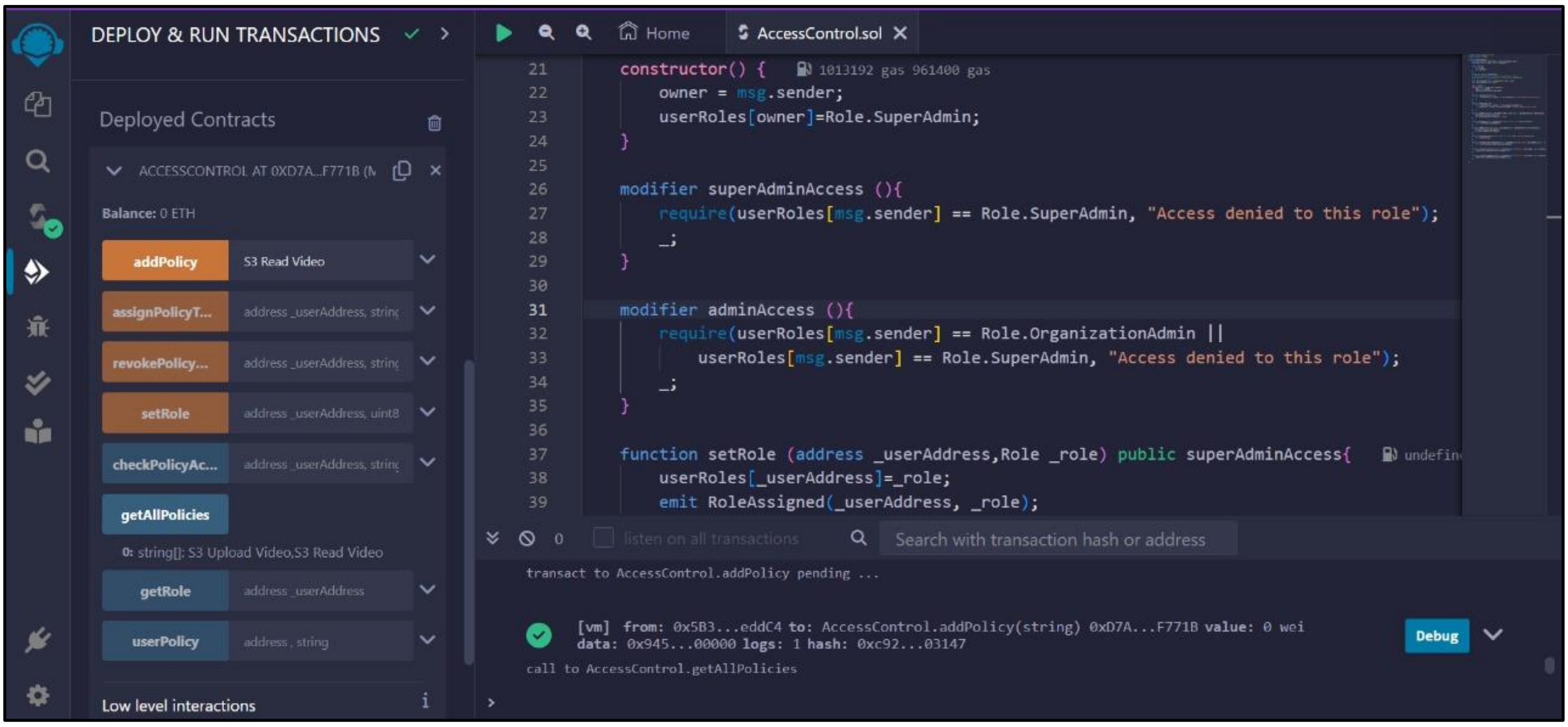

***Figure 6:*** *Creating Policy in Blockchain*

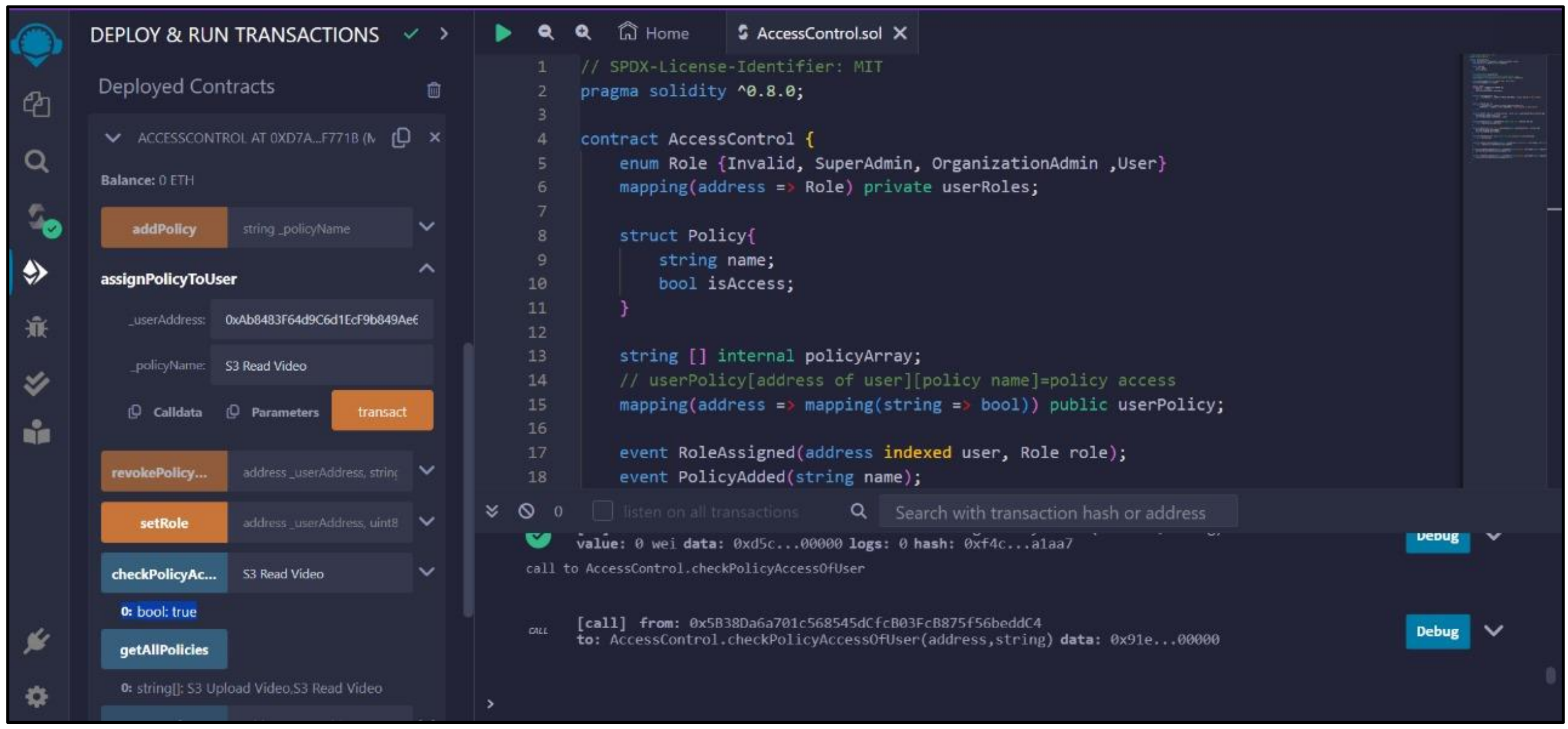

***Figure 7:*** *Granting Permission to a user via unique userId and functionalityName*

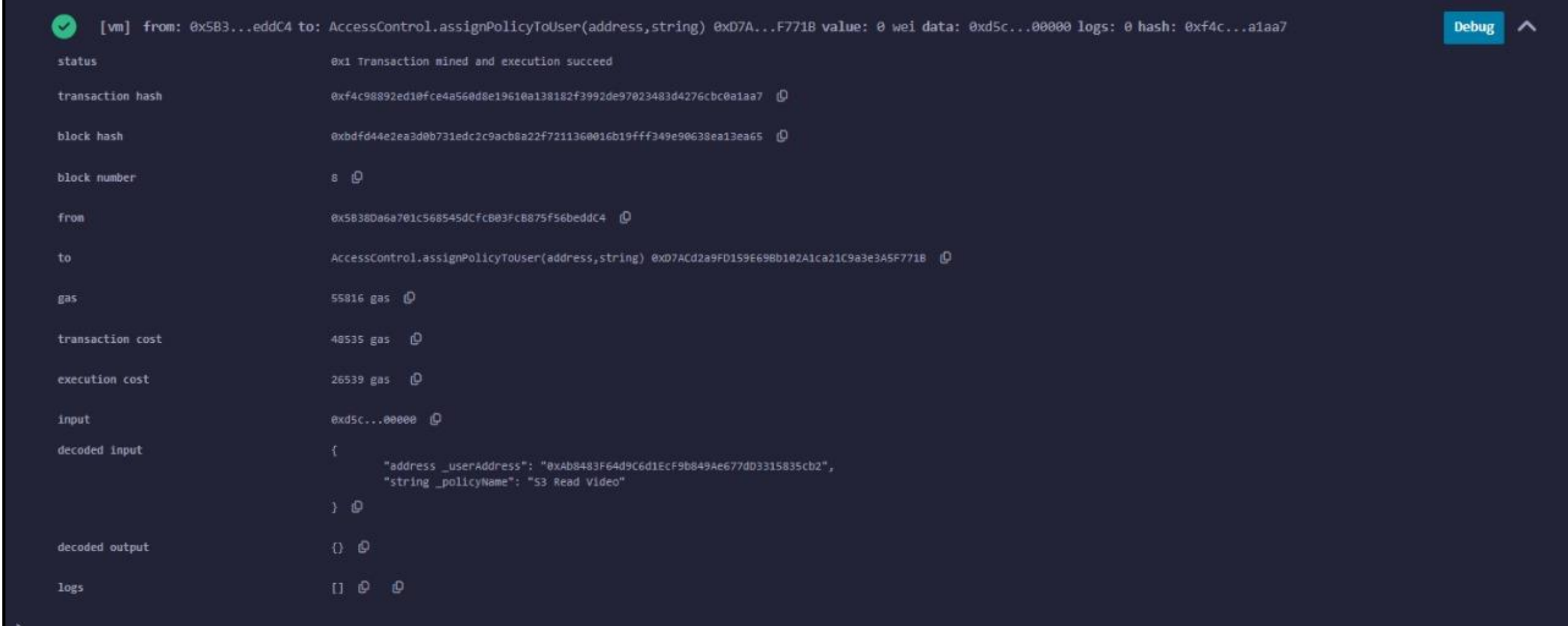

***Figure 8:*** *Successful Transaction*

## 6.2 Overcoming Third-Party Dependency in Data Integrity Assurance

This architecture uses digital certificates to maintain the integrity of high-frequency data transactions occurring in the cloud environment. However, the conventional working of digital certificates relies on a third party i.e. Certifying Authority (CA) for the issuance and validation of certificates. The involvement of such external entities increases the risk of unauthorized access and malicious activities thus, leading to potential compromise of security concerns. We tackle this issue by using smart contracts and encoding them with predefined rules. Their cryptographic underpinning ensures that even minimal tampering in the data can be easily identified. Hence this automates the process of data transfer between different parties in a secure way contributing to a resilient system.

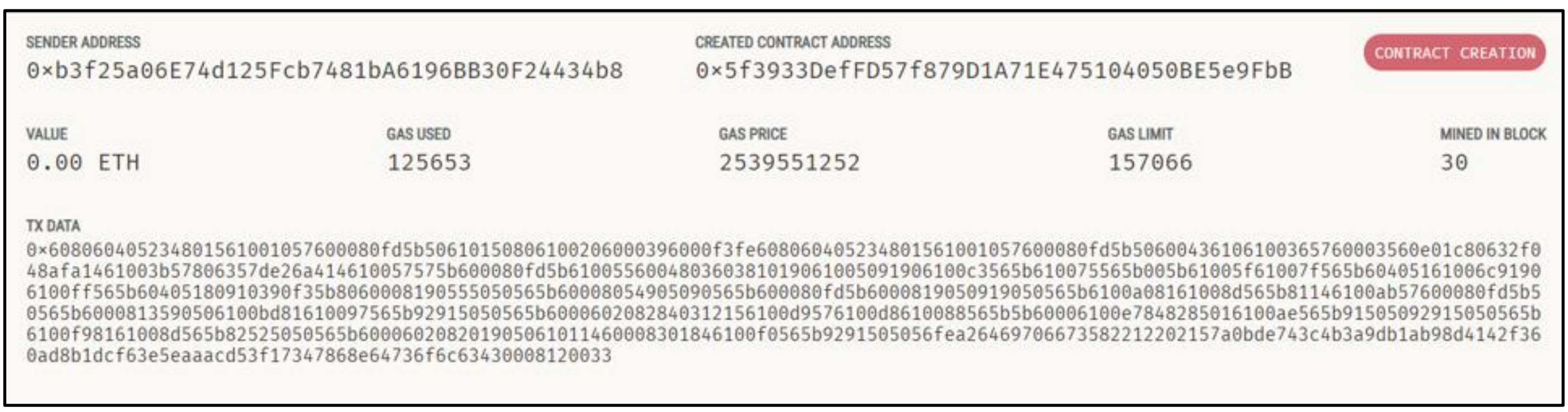

***Figure 9:*** *Cryptographic Hashing*

| BLOCK | MINED ON | GAS USED | |
|---|---|---|---|
| BLOCK 30 | MINED ON 2024-03-02 11:18:15 | GAS USED 125653 | 1 TRANSACTION |
| BLOCK 29 | MINED ON 2024-03-02 11:14:19 | GAS USED 125653 | 1 TRANSACTION |
| BLOCK 28 | MINED ON 2024-03-02 10:35:35 | GAS USED 26602 | 1 TRANSACTION |

***Figure 10:*** *Audit Trail of different Blocks*

## 6.3 Addressing Processing Fees and Feasibility in Blockchain-Cloud Frameworks

Integration of blockchain with cloud computing, while enhancing security, may introduce network congestion and slower transaction processing speeds. Thus, striking a balance between block size and transaction throughput is crucial in optimizing the performance. Block size is critical in determining the computation speed and storage requirements. This is because a large block size can accommodate more transactions but may lead to slower validation speed and increased storage requirements. Smaller blocks enhance transaction processing speed but limit scalability. Hence, the proposed framework resolves this issue by storing the fixed size of cryptographic hashes of the digital certificate on the blockchain which is interlinked with the off-chain repositories. Also, if all cloud data is stored on blockchain this will increase the processing fees and will reduce the feasibility of the proposed architecture. So, this scheme of on-chain verification and off-chain storage will also be advantageous for organizations to adopt this solution and connect their existing database systems with blockchain instead of migrating whole data to blockchain networks.

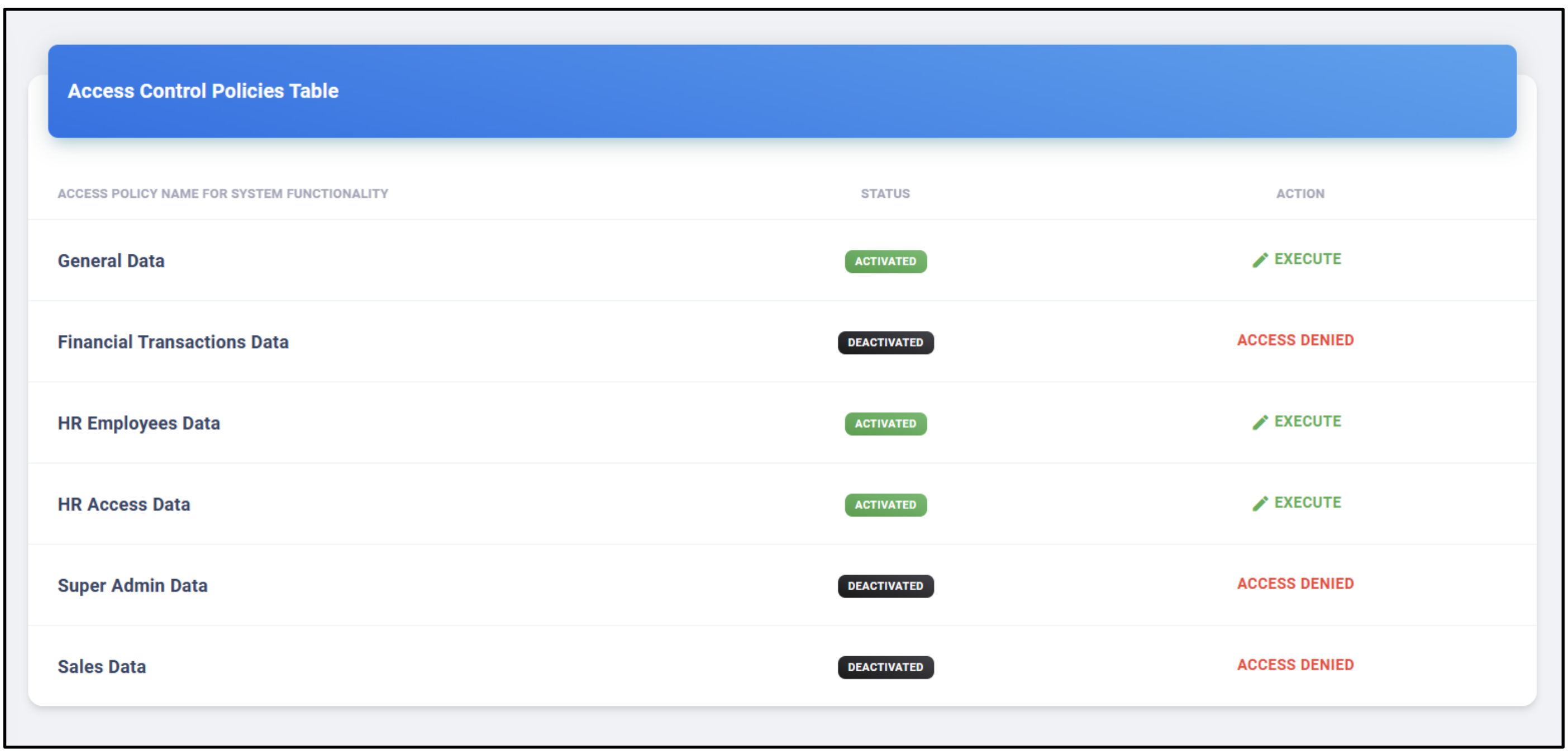

Access Control Policies Table

| ACCESS POLICY NAME FOR SYSTEM FUNCTIONALITY | STATUS | ACTION |
|---|---|---|
| General Data | ACTIVATED | EXECUTE |
| Financial Transactions Data | DEACTIVATED | ACCESS DENIED |
| HR Employees Data | ACTIVATED | EXECUTE |
| HR Access Data | ACTIVATED | EXECUTE |
| Super Admin Data | DEACTIVATED | ACCESS DENIED |
| Sales Data | DEACTIVATED | ACCESS DENIED |

***Figure 11:*** *Front end depicting functionalities an individual can execute*

## VII. Conclusion

The proposed blockchain-based architecture stands as a resilient solution and lays the groundwork for Trust Management in the realm of the public Cloud environment. As it addresses key challenges such as vulnerabilities in access control, third-party dependency in data integrity assurance, and processing fees in blockchain-cloud integration. Moreover, this combination provides an appealing solution for organizations who are seeking enhanced security with their current cloud infrastructure. The framework's design minimizes processing fees by implementing an optimized methodology linking block size and transaction throughput. In essence, this holistic solution will create a strategic pathway for organizations. Thus, organizations will seamlessly integrate blockchain components with their existing cloud infrastructure without requiring a complete migration of data to the blockchain network.